\documentclass[11pt,singlecolumn,showpacs,preprintnumbers,amsmath,amssymb,prd]{revtex4}

\usepackage{latexsym}
\usepackage{amssymb}
\usepackage{epsfig,amsmath,graphics}
\usepackage{epstopdf}
\usepackage{verbatim}
\usepackage{wasysym}
\usepackage{feynmp-auto}
\usepackage{xcolor}

\newcommand{\OO}{\mathcal{O}}

\newcommand{\ket}[1]{\ensuremath{\left|#1\right>}}
\newcommand{\bra}[1]{\ensuremath{\left<#1\right|}}

\begin{document}


\title{Particle Probes with Superradiant Pulsars}

\author{David E. Kaplan}
\affiliation{Department of Physics and Astronomy, The Johns Hopkins University, 3400 N. Charles Street, Baltimore, MD 21218, USA}


\author{Surjeet Rajendran}
\affiliation{Department of Physics and Astronomy, The Johns Hopkins University, 3400 N. Charles Street, Baltimore, MD 21218, USA}

\author{Paul Riggins}
\affiliation{Berkeley Center for Theoretical Physics, Department of Physics,
University of California, Berkeley, CA 94720, USA}

\date{\today}

\begin{abstract}
  We demonstrate that rotational superradiance can be efficient in millisecond pulsars. Measurements from the two fastest known pulsars PSR J1748-2446ad and PSR B1937+21 can place bounds on bosons with masses below $10^{-11}~\text{eV}$. The bounds are maximally good at masses corresponding to the rotation rate of the star, where scalar interactions that mediate forces $ \sim 10^6$ times weaker than gravity are ruled out, exceeding existing fifth force constraints by 3 orders of magnitude. For certain neutron star equations of state, these measurements would also constrain the QCD axion  with masses between $5\times10^{-13}$ and $3 \times 10^{-12}~\text{eV}$.
   Despite the ability of most neutron star equations of state to support frequencies as high as $ \sim 1500~\text{Hz}$, the observed absence of pulsars above $ \sim 700~\text{Hz}$ could be due to the 
   existence of a new particle of mass $ \sim 2\pi\times 1500-3000~\text{Hz}$ or $\sim 10^{-11}$ eV with a Yukawa coupling to nucleons.
 \end{abstract}

\maketitle
\tableofcontents

\section{Introduction}
\label{Sec:Introduction}

Ultra-light bosonic particles that interact with ultra-low couplings to the standard model are an interesting target to search for new physics. Such particles emerge in a variety of contexts. They are prime dark matter candidates \cite{Hu:2000ke} or can act as mediators between the standard model and the dark sector. They may also emerge naturally in the context of ultra-weakly coupled gauge theories or in cosmological relaxation scenarios where the evolution of the universe can lead to ultra-light particles in a sufficiently old universe \cite{Graham:2015cka, Graham:2019bfu, Graham:2017hfr}. Currently, the strongest reliable constraints on the existence of such particles with mass below $\sim$ eV are placed by direct laboratory searches in Cavendish experiments \cite{Adelberger:2003zx}.  A more sensitive way to search for such particles was suggested in  \cite{Arvanitaki:2009fg, Arvanitaki:2010sy}, using the superradiance instability of black holes. It is well known that rotating black holes can lose their angular momentum through excitation of particles whose masses are close to the rotational frequency of the black hole.  The authors of \cite{Arvanitaki:2009fg, Arvanitaki:2010sy} point out that the rotational frequency of extremal astrophysical black holes can be close to the masses of interesting light particles, such as the QCD axion. Constraints on these particles can be placed through observations of rotating black holes. Alternately, gaps in the spectrum of rotating black holes can be used to discover particles whose masses correspond to that rotational frequency.
Extremal, stellar mass black holes most effectively probe mass scales $\sim 10^{-9}$ eV (corresponding to rotation rates $\sim$ 100 kHz).
More recently, extremal black holes of masses $\sim 10~M_\odot$ have been used to place superradiant constraints on lighter particles \cite{Arvanitaki2015,Baryakhtar2017}, including masses comparable to what we will study here.

 The applicability of this interesting idea is  limited by difficulties in directly measuring the rotation rate of black holes \cite{BlackHoleSpin}.  The rotation rate is not directly measured - it is instead inferred either by models of the jets emerging from the black hole or through fits of the spectrum of accretion disk emissions. The superradiant instability is a strong function of the rotation rate of the black hole: while a nearly extremal black hole would have a rapid superradiant instability, a black hole that spins only $\sim$ 30 - 50 percent slower would not be significantly affected by the superradiant instability.
In addition to these observational difficulties, there are also theoretical uncertainties. The calculations of  \cite{Arvanitaki:2009fg, Arvanitaki:2010sy, Arvanitaki2015, Baryakhtar2017} assume that the geometry of the black hole is described by the Kerr solution without any matter sources just outside the event horizon.
While this is a conventional assumption, it is well known that if all of the conventional assumptions about black hole physics are correct, there cannot be a solution to the black hole information problem \cite{Mathur:2009hf}. The existence of a singular firewall just outside the horizon of the black hole is a plausible resolution to this problem \cite{Almheiri:2012rt}. Recently, it has been shown that such firewall solutions are in fact compatible with General Relativity \cite{Kaplan:2018dqx}. If these firewalls exist they can change the boundary condition just outside the horizon, sourcing deviations away from the axisymmetric assumptions made in the calculations of \cite{Arvanitaki:2009fg, Arvanitaki:2010sy}.  Specifically, these deviations can cause mixing between superradiant and absorptive modes, potentially dampening the growth of such modes.

It is thus interesting to ask if the superradiance instability can be effective in other astrophysical objects whose properties are better  understood observationally. Superradiance as a general instability of rotating systems was discovered well before its application to the rotation of black holes \cite{Zeldovich}. The only aspect of black hole physics necessary for the existence of this instability is the absorption  provided by the black hole horizon for the particle \cite{arXiv:0909.2317, gr-qc/9803033, Zeldovich}. In this paper, we argue that these conditions can also be satisfied for another class of extremal, rotating objects, namely, millisecond pulsars. Unlike a black hole, the gravitational forces exerted by such a pulsar are not strong enough to create an absorptive region for the particle. However, such an absorptive region can be provided by non-gravitational interactions of the particle with the stellar medium.

We show that an absorptive coupling to light particles can be sufficient to slow down the rotation rates of millisecond pulsars provided the particles have masses $\sim 10^{-11}$ eV ($\sim$ kHz).
Unlike black holes, millisecond pulsars are easily discovered through electromagnetic signals. Further, in contrast to measurements of black hole rotation, the frequencies of millisecond pulsars are the most precisely known numbers in astrophysics.  Since the composition of the pulsar is known, it is also possible to reliably estimate deviations from axisymmetry and show that the growth of the superradiant mode is not damped by mixing with absorptive modes.  Consequently, the existence of these objects can be used to place a robust bound on particles of mass $\sim 10^{-11}$ eV that couple sufficiently strongly with the stellar medium.
While the possibility of using superradiant pulsars to constrain such particles has been discussed before \cite{Cardoso:2015zqa,Cardoso:2017kgn,Day:2019bbh}, only the stellar conductivity has been concretely considered as a dissipation mechanism. Perhaps more significantly, the effects of mixing with absorptive modes have not yet been carefully considered, though they are necessary to place realistic constraints.

The rest of the paper is organized as follows. In Section \ref{Sec:Superradiance}, we review the phenomenon of superradiance and show that it is applicable to a wide variety of rotating systems. The formalism necessary to estimate the superradiance rates of particles coupled to the stellar medium is developed in Section \ref{Sec:NeutronStarSuperradiance}. After examining the feasibility of superradiance in real astrophysical environments, bounds on particle models are placed in Section \ref{Sec:Constraints}, and we conclude in Section V.

\section{Superradiance}
\label{Sec:Superradiance}

A rotating body can spin down by emitting light degrees of freedom. This radiative emission requires two conditions. First, the degrees of freedom must be light enough so that there is phase space available for the process. Second, there must be a non-zero matrix element between the rotating medium and the light degree of freedom. Consider an isolated, axi-symmetric rotating object. There is phase space available for this object to spin down, for example, through the emission of a photon or other suitably light degree of freedom. The emitted particle needs to carry angular momentum away from the rotating object - in an inertial reference frame centered on the rotating body, the emitted particle will have a non-zero azimuthal angular quantum number. But, when the rotating body is axisymmetric, this particle cannot be emitted since the coupling between the rotating body and the kinematically accessible, angular momentum carrying mode vanishes due to the axisymmetry. While the leading order process is forbidden, there can be higher order processes. For example, if the rotating object has soft deformations ({\it e.g.} phonons), these deformations break the axisymmetry and can couple to the kinematically allowed emissive  mode. Thus, the body can spin down by simultaneously producing the light degree of freedom while sourcing soft deformations on itself (which are eventually damped away through other dissipative effects). Clearly, when this process can occur in a rotating body, it will also be possible for the body to absorb the light particle when it is non-rotating: in this case, the absorption leads to deformations of the body and an increase in its angular momentum. Hence, the existence of absorption signals the possibility of superradiant emission when the emission is kinematically allowed.

This description of superradiance and its subsequent effects can be captured by the following set of equations. Consider an object that is coupled to a light degree of freedom $\Psi$ with mass $\mu$. The interactions of $\Psi$ with the object will induce an absorptive term in its equation of motion:
\begin{align}
\square \Psi + \mu^2 \Psi + C v^{\alpha}\nabla_{\alpha}\Psi + V_{eff}\left(\Psi\right) & = 0
\label{Eqn:Absorption}
\end{align}
where $C$ is the absorption coefficient, $v^{\alpha}$ the four velocity of the system and $V_{eff}\left(\Psi\right)$ is any other potential that dictates the motion of $\Psi$. This expression is  the covariant generalization of the familiar equation for absorption in the rest frame of the system
\begin{align}
\square \Psi + \mu^2 \Psi + C  \dot{\Psi} + V_{eff}\left(\Psi\right) & = 0
\end{align}
since in the rest frame $v^{\alpha} = \left(1, 0, 0, 0\right)$, were we assume $C$ has at most a weak dependence on $v^{\alpha}$. Owing to absorption, an initial amplitude of $\Psi$ exposed to this system will decay exponentially as $e^{-\frac{C}{2} t}$.

Let the system rotate with frequency $\Omega$. Choose spherical coordinates $\left(t, r, \theta, \phi\right)$ centered around the axis of rotation. In these coordinates, $v^{\alpha} = \left(1, 0, 0, \Omega \, r \sin \theta\right) + \OO\left(\left(\Omega r\right)^2\right)$. Now, consider the equation of motion for a specific angular momentum mode of $\Psi$. These are of the form $\tilde{\Psi}\left(r, \theta \right) e^{-i E t}e^{i m \phi}$ where $E$ is the energy and  $m$ the azimuthal angular quantum number of the mode. For a non-relativistic mode, the energy $E$ is dominated by the rest mass $\mu$ of the particle. For this mode,
 the absorption term in 
the equation of motion takes the form: 
\begin{equation}
C v^{\alpha}\nabla_{\alpha}\Psi \rightarrow - i C  \left( \mu - m \, \Omega\right)\tilde{\Psi}\left(r, \theta\right)
\label{Eqn:EffectiveSuperradiance}
\end{equation}
 Thus in the equations of motion, for sufficiently large $m \, \Omega$, $\mu - m \, \Omega < 0$ and  the term $C \left(\mu - m \, \Omega\right)$ flips sign. This converts the absorptive term into an emissive term leading to exponential growth $\propto e^{\frac{C}{2} t}$ of $\Psi$. This exponential growth is indicative of emission of $\Psi$ by the system, leading to energy loss from the system through the decay of its rotational energy.

Superradiance is thus a general instability of rotating systems that leads to the decay of the rotational energy in the system. In the next section, we will show that this instability can be very efficient in compact, rapidly rotating systems such as neutron stars if the stellar medium couples to light particles whose masses are order the rotation rate $\Omega$ of the star.

\section{Superradiance in Neutron Stars}
\label{Sec:NeutronStarSuperradiance}
Superradiance results in the conversion of rotational kinetic energy into excitations of certain angular momentum modes of particles coupled to the rotating medium. The rate of superradiance is governed by the operator in \eqref{Eqn:EffectiveSuperradiance}, where the superradiant term appears in the same form as an absorptive term, but has the opposite sign. Consequently, much like absorption, the rate of superradiance is proportional to the occupation number of the concerned mode. In this way, superradiance can be thought of as a form of stimulated emission. The amplitude of a bosonic superradiant mode will grow at a rate proportional to its occupation number, resulting in exponential amplification of the mode. This exponential increase in the amplitude will lead to exponential energy loss from the rotating system.  On the other hand, a superradiant fermionic mode will not lead to such an exponential energy loss since Pauli exclusion leads to the saturation of the mode's amplitude once it acquires one particle, thereafter shutting off the superradiant channel. This exponential growth  occurs in the region where the mode overlaps with the rotating medium. It is only in this region that  \eqref{Eqn:EffectiveSuperradiance} contributes to the equation of motion of the mode. The efficiency of superradiant energy loss thus depends strongly on the overlap between the rotating medium and the superradiant mode.

\begin{figure}
\begin{center}
\includegraphics[width = 4.5 in]{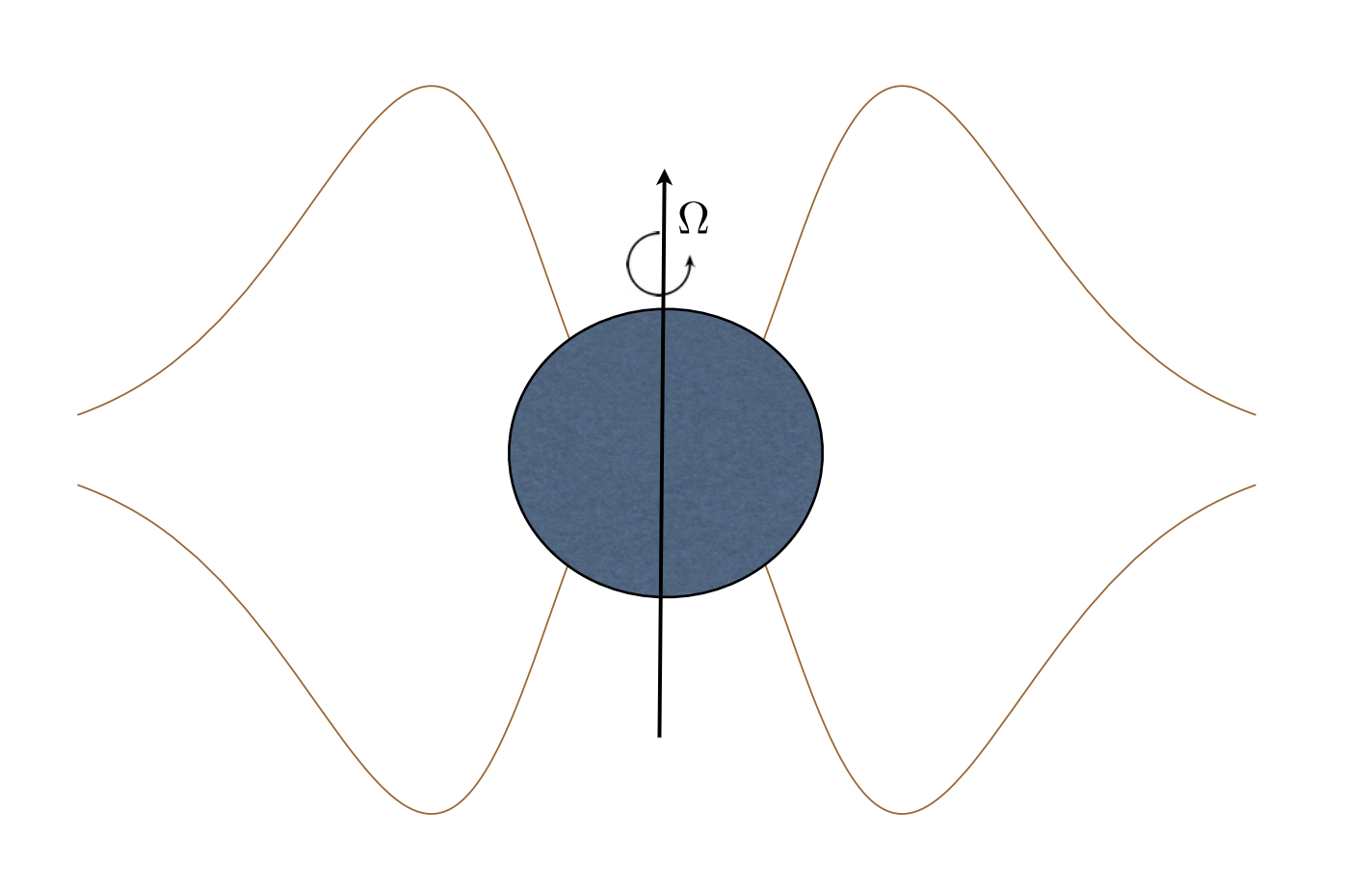}
\caption{ \label{Fig:modefig} A state with non-zero azimuthal angular momentum bound gravitationally to a neutron star. The star spins with frequency $\Omega$. Superradiant growth occurs in the region where the mode overlaps with the star. }
\end{center}
\end{figure}

The rotational angular momentum modes of a light bosonic particle bound gravitationally to a spinning neutron star satisfy the characteristics discussed in the above paragraphs to act as an efficient superradiant conduit (see Figure~\ref{Fig:modefig}). These modes are solutions to Equation \eqref{Eqn:Absorption} where the potential $V_{eff}$ is given by the gravitational interaction energy between the star and the bound particle. The non-relativistic limit of this equation is obtained by decomposing the field $\Psi$ in the form $e^{-i \mu t} \psi_{nlm}\left(t, r, \theta, \phi\right) $ and dropping time derivates of order $\frac{\dot{\psi}_{nlm}}{\mu}$ and higher, yielding \label{NonRelativisticApproximation}
\begin{align}
i \dot{\psi}_{nlm} =  - \frac{1}{2 \mu} \left( \nabla^2 \psi_{nlm} \right)  - \frac{G \, M \, \mu}{r} \psi_{nlm} + i \frac{C \left( \mu - m \, \Omega \right)}{2 \, \mu} \psi_{nlm}
\label{Eqn:BoundEquation}
\end{align}
Without the absorptive term $\propto C$ on the right-hand side, equation \eqref{Eqn:BoundEquation} is the Schrodinger equation describing a mode $\psi_{nlm}$ with radial quantum number $n$, total angular momentum $l$ and azimuthal angular momentum $m$, moving in the gravitational potential of a star of mass $M$. Assuming the star to be spherically symmetric (we will discuss the effects of deviations away from spherical and axisymmetry in section \ref{Sec:Stability}), the modes $\psi$ are the usual Hydrogenic wavefunctions with real eigenenergies that correspond to the bound state energy. These modes are localized around the ``Bohr'' radius $\sim  \frac{n^2}{\alpha_g  \, \mu }$ where the gravitational ``fine structure'' constant $\alpha_g = G M \mu$ (see Figure~\ref{Fig:modefig}).

The absorptive term in \eqref{Eqn:BoundEquation} is  non-Hermitian and leads to these modes developing imaginary eigenenergies, indicating growth or decay of the mode. To estimate this imaginary part, note  that the absorptive term is non-zero only in the interior of the star. For physically relevant neutron stars, it will turn out that the mass of the particles that can undergo superradiance are such that the Bohr radius of the mode is much bigger than the radius $R$ of the star. The absorptive term only affects a small part of the wavefunction and its effects can be estimated using perturbation theory.   The imaginary part of the energy shift caused by this perturbation is

\begin{equation}
 \frac{\Gamma_{nlm}}{2} = \langle \psi_{nlm} |\frac{C}{2} \frac{\left(\mu - m\, \Omega\right)}{\mu} |\psi_{nlm} \rangle \equiv \frac{C_{nlm}}{2} \frac{\left(\mu - m\, \Omega\right)}{\mu}
\label{Eqn:rate}
\end{equation}
Physically, this corresponds to the expectation that the mode can only grow/decay inside the star and hence the growth rate is proportional to the probability of finding the particle in that region (see Figure~\ref{Fig:modefig}). For $ \mu > m \, \Omega$, the imaginary part is positive, leading to absorption and  exponential damping of the mode. When  $ \mu <  m \, \Omega$, the imaginary part is negative, leading to emission and exponential amplification of the mode. In both cases, the rate of absorption/emission is given by \eqref{Eqn:rate}.

Using the rate \eqref{Eqn:rate}, it is easy to see that efficient superradiance requires two conditions. First, equation \eqref{Eqn:rate} is a strong function of the size $R$ of the star since the probability of finding the particle within the star depends upon its size. This size is however limited by the rotational frequency $\Omega$ of the star, since relativistic considerations require that $\Omega R < 1$. Consequently, superradiance is most efficient in objects that are close to satisfying this bound. Second,  equation \eqref{Eqn:rate} is also a strong function of the angular momentum $m$ required to achieve the superradiance condition. This is because modes with high angular momentum are localized away from the star, leading to a suppressed probability of finding the particle inside the star. Consequently, superradiance is most effective when the mass of the particle $\mu$ is close to $\Omega$. In this case, the superradiance condition $\mu - m \, \Omega < 0$ will be satisfied by low angular momentum modes $m \approxeq 1$. When $\mu \gg \Omega$, the superradiance condition will only be satisfied by modes with very high angular momentum. But, these modes are at Bohr radii ($\propto n^2 \gtrsim m^2$) far from the star leading to a highly suppressed overlap with the star and hence a suppressed superradiance rate. Similarly, when $\mu \ll \Omega$, even though the superradiance condition is satisfied by many low lying modes, the Bohr radius of the mode ($\propto \mu^{-2}$) is also far from the star leading to  suppressed overlap.

These considerations suggest that superradiance could be efficient in millisecond pulsars, due to the large angular momentum.  A typical  millisecond pulsar has a radius $R \sim 10 - 15$ km, with rotational frequency $\Omega \sim 2 \pi(1\, \textrm{kHz})$, close to saturating the extremality bound $\Omega R < 1$.  The very existence of such pulsars should  constrain the existence of particles with  masses $\mu \sim \Omega \sim 10^{-11}$ eV that couple sufficiently strongly to the stellar medium. We pursue this question in the rest of the paper, starting with section \ref{Sec:Rate} where we estimate the superradiance rate for particles that are coupled to the stellar medium.

\subsection{Rate}
\label{Sec:Rate}
In this section, we estimate the superradiance rate \eqref{Eqn:rate} for the states $|\psi_{nlm}\rangle$ of a scalar field $\Psi$ that are bound gravitationally to a neutron star. This rate, up to the kinematic ratios in \eqref{Eqn:rate}, is the absorption rate $C_{nlm}$ of the mode $|\psi_{nlm}\rangle$ in the stellar medium when the medium is at rest. Energy and angular momentum have to be conserved in this absorption process. This requires the excitation of inelastic degrees of freedom in the stellar medium, in addition to energy and angular momentum being transferred to the bulk stellar rotation. The energies of these inelastic degrees of freedom have to be comparable to the energy of $|\psi_{nlm}\rangle$ and are therefore $\sim \Omega$. In the stellar medium, these light degrees of freedom can be phonon modes of the neutrons or other low frequency oscillations, for instance. For simplicity, we will compute the superradiant emission of $\Psi$ when it has scalar interactions with the stellar medium.

\subsubsection{Scalar Absorption Rate}
\label{sec:ScalarAbsorptionRate}

We now turn to the main operator of interest for this paper, the neutron Yukawa interaction
\begin{equation}
\epsilon\, \Psi\, \overline{n}\, n.
\label{Eqn:ScalarLagrangian}
\end{equation}
We may use this to probe any new ultralight scalar or CP-violating pseudoscalar. Depending on the neutron star equation of state, this may even include the QCD axion: many neutron star equation of states predict a pseudoscalar condensate in the star, throughout $\mathcal{O}(1)$ of the star's mass \cite{Lattimer2001}. In this phase we expect $\theta_\text{eff} \sim 1$, in which case the QCD axion obtains a neutron coupling $\epsilon \sim \theta_\text{eff}\, m_n / f_a \sim m_n / f_a$, which is large enough to probe new regions of parameter space.

In order to estimate the scalar-phonon conversion rate, we begin with 1D toy model which we believe captures the essence of the process, and extrapolate to 3D at the end. Take a string of $N$ neutrons spread over a length $R$. The absorption of $\Psi$ results in phonon excitations of the string. Let us enumerate these phonon excitations $|k\rangle$. The string consists of $N$ neutrons and we assume that these neutrons have nearest neighbor interactions. Since we ultimately want to model a neutron star, we will take the string to contain a nuclear density of neutrons with the strength of nearest neighbor couplings set by the QCD scale.  For small displacements, these nearest neighbor interactions will be harmonic and the states $|k\rangle$ correspond to phonon excitations of the string. The neutrons and the $\Psi$ particles are non-relativistic throughout this process and are modelled with the non-relativistic ``free'' hamiltonian (\emph{i.e.} neglecting \eqref{Eqn:ScalarLagrangian}):
\begin{align}
H_{F} & =   \frac{p_{\Psi}^2}{2 \, \mu}  + \sum_{j = 1}^{N} \frac{p_j^2}{2 \, m_n} + \frac{1}{2} m_n \omega^2 \left(\delta x_j - \delta x_{j-1} \right)^2
\label{Eqn:NearestHamiltonian}
\end{align}
where $\delta x_j$ is the displacement from the equillibrium position $x_j^{0}$ of the $j^{th}$ neutron, $p_j$ the corresponding conjugate momemtum, $m_n$ the mass of the neutron, $\omega \sim \Lambda_{QCD}$ the strength of the nearest neighbor interaction and $p_{\Psi}$ the conjugate momentum of $\Psi$.

The Hamiltonian \eqref{Eqn:NearestHamiltonian} can be diagonalized through a coordinate transformation given by 
\begin{align}
\delta x_j &= \sum_{s = 1}^{N} y_{js} Y_s
\end{align}
where $y_{js}$ are the normalized wavefunctions, after which the Hamiltonian becomes
\begin{align}
H_{F} & =  \frac{p_{\Psi}^2}{2 \, \mu}  +\sum_{j=1}^{N} \frac{|q_j|^2}{2 \, m_n} + \frac{1}{2} m_n \omega_j^2 |Y_j|^2
\label{Eqn:FreeHamiltonian}
\end{align}
where $q_j$ are the conjugate momenta of the coordinates $Y_j$ and the frequencies $\omega_j \sim \frac{j}{N} \, \omega$ for $j\ll N$. 
The normalized wavefunctions are given approximately by $y_{js} \sim N^{-1/2}\exp(i2 \pi js/N)$ . Notice the normalization suppression by $\sqrt{N}$ due to the participation by all $N$ neutrons in the oscillation. In terms of the new phonon coordinates $Y_j$ and these wave functions $y_{js}$, the position $x_j$ of the $j^{th}$ neutron is given by
\begin{equation}
x_j  =  x^{0}_j + \delta x_j = x^{0}_j +  \sum_{s = 1}^{N} y_{js} Y_s
\label{Eqn:xjexpansion}
\end{equation}

The Hamiltonian \eqref{Eqn:FreeHamiltonian} describes $N$ free harmonic oscillators with frequencies between $\frac{\omega}{N}$ and $\omega$. These correspond to sound waves (phonons) in the one dimensional string of neutrons, with quantized frequencies. The quantization unit for the frequencies is set by the number of neutrons in the string. In the one dimensional example, this number is directly proportional to the length $R$ of the string, resulting in quantization set by the physical size of the system as one might expect for a sound wave. The eigenstates $|k\rangle$ of this system are given by $\prod_{s = 1}^{N} |k_s\rangle$ where $|k_s\rangle$ is an eigenstate of the free harmonic oscillator with frequency $\omega_s$ and occupation number $k_s$.

We now have  a description of the string. Before proceeding with the computation of the  absorption coefficient, we must also model the interaction of the scalar field $\Psi$ with the neutrons. The perturbation \eqref{Eqn:ScalarLagrangian} caused by $\Psi$ is a shift to the mass of the neutron. The resulting total hamiltonian that also includes these interactions is
\begin{align}
H & = H_F + \epsilon \sum_{j = 1}^{N} \Psi\left(x_j\right)
\label{Eqn:ScalarHamiltonian}
\end{align}
We will represent  $\Psi$ in terms of its creation and annihilation operators ($a_{\Psi}$ and $a_{\Psi}^{\dagger}$ respectively). This is necessary since absorption requires operators that can destroy particle number.  In this representation, $\Psi$ is given by
\begin{equation}
\Psi \left(x_j \right) = \int \frac{dp}{ \left(2 \pi\right)} \, \frac{1}{\sqrt{2 E_p}} \, \left(a_{\Psi,p}^{\dagger} e^{-i px_j} \,  + \,  a_{\Psi,p} e^{i px_j} \right)
\end{equation}
where $E_p$ is the energy of the state of momentum $p$. The $\Psi$ particles absorbed by the string of neutrons are also non-relativistic and hence $E_p \sim \mu$.


The above states are the eigenstates in the free theory. Of course, we need these states in the full interacting theory. These can be calculated using the Lippmann-Schwinger equation,
\begin{align}
| k,  \psi_{nlm} \rangle_\text{int}  =   \left(1 + G_{+} \epsilon \sum_{j = 1}^{N} \Psi \left(x_j\right) +  G_{+}  \left(\epsilon \sum_{j = 1}^{N} \Psi \left(x_j\right)\right) G_{+} \left(\epsilon \sum_{l = 1}^{N} \Psi \left(x_l\right)\right)  + \dots \right) | k, \psi_{nlm} \rangle
\label{Eqn:LippmannSchwinger}
\end{align}
where $G_{+}$ is the retarded Green's function of the free phonon Hamiltonian $H_F$ with energy $E$ equal to the total initial energy of the system. Formally, $G_{+}$ is obtained by inverting $E - H_{F}$. Using these states and a form of the optical theorem, the absorption rate is
\begin{align}
  \label{Eqn:RateExpansion}
C_{nlm} = \text{Im}\left( \langle  k, \psi_{nlm} |  \left(\left(\epsilon \sum_{j = 1}^{N} \Psi \left(x_j\right)\right) +   \left(\epsilon \sum_{j = 1}^{N} \Psi \left(x_j\right)\right) G_{+} \left(\epsilon \sum_{l = 1}^{N} \Psi \left(x_l\right)\right)  + \dots \right)  | k, \psi_{nlm} \rangle\right)
\end{align}
The second term in  \eqref{Eqn:RateExpansion}  is the  lowest order (in $\epsilon$) term that can give rise to imaginary coefficients. This term allows for absorption of $\Psi$ and excitation of phonons, followed by propagation of the excited phonon states and then subsequent re-emission of $\Psi$ and de-excitation of phonons. The Green's function $G_{+}$ develops poles from the propagation of the on-shell,  excited phonon states. These poles are regulated by the width $\Gamma$ of the intermediate states, yielding imaginary coefficients.

More concretely, the imaginary part  is
\begin{align}
C_{nlm} = \text{Im}\left( \langle  k, \psi_{nlm} |   \left(\epsilon \sum_{j = 1}^{N} \Psi \left(x_j\right)\right) G_{+} \left(\epsilon \sum_{l = 1}^{N} \Psi \left(x_l\right) \right)  | k, \psi_{nlm} \rangle\right)  + \OO\left(\epsilon^3\right)
\label{Eqn:ImaginaryPart}
\end{align}
Inserting a complete set of intermediate phonon states $\sum_{k'} |k'\rangle \langle k'|$ into \eqref{Eqn:ImaginaryPart}, we get
\begin{align}
\langle  k, \psi_{nlm} |\,   \left(\epsilon \sum_{j = 1}^{N} \Psi \left(x_j\right)\right) \sum_{k'} |k'\rangle \langle k'| G_{+} \sum_{k''} |k''\rangle \langle k''|\left(\epsilon \sum_{l = 1}^{N} \Psi \left(x_l\right)\right)\,| k, \psi_{nlm} \rangle
\label{Eqn:MiddleStepTwo}
\end{align}
The propagator $ \langle k'| G_{+}  |k''\rangle$  of the intermediate phonon states in \eqref{Eqn:MiddleStepTwo} is obtained by inverting the free phonon hamiltonian $E - H_F$ and is
\begin{align}
\langle k'| G_{+}  |k''\rangle &= \frac{\delta_{k'k''}}{E - E_{k'} + i \, \Gamma_{k'}}
\label{Eqn:PhononPropagator}
\end{align}
The parameters $E_{k'}$ and $\Gamma_{k'}$ in this expression are of course the energy and decay rate of the state $|k'\rangle$. Using \eqref{Eqn:PhononPropagator},  \eqref{Eqn:MiddleStepTwo} is equal to
\begin{align}
\sum_{k'}\langle  k, \psi_{nlm} |\,  \left(\epsilon \sum_{j = 1}^{N} \Psi \left(x_j\right)\right)  |k'\rangle  \frac{1}{E - E_{k'} + i \, \Gamma_{k'}}  \langle k'|\left(\epsilon \sum_{l = 1}^{N} \Psi \left(x_l\right)\right)\,| k, \psi_{nlm} \rangle
\label{Eqn:MiddleStepThree}
\end{align}

The next task is to compute the transition elements in \eqref{Eqn:MiddleStepThree} that lead to the excitation of phonon modes. In the problem of interest, $\Psi$ is a light field, with $\mu \lesssim \omega_1$. We expect the dominant contribution to the transition element is the excitation of the lowest phonon states while leaving the other states unperturbed. $|k'\rangle$ is therefore of the form $|k_1 + 1\rangle \otimes \prod_{s = 2}^{N} |k_s\rangle$, and so we only need the one-phonon contribution from the interaction potential. Taylor expanding the scalar operator to first order about the neutron equilibrium positions, we find
\begin{align}
\epsilon \sum_{j = 1}^{N} \Psi \left(x_j\right) &\approx \epsilon \sum_{j = 1}^{N} \left( \Psi (x _j^0) + \left.\frac{\partial \Psi}{\partial x}\right|_{x_j^0}\delta x_j \right)
\label{Eqn:OperatorExpansion}
\end{align}
The first term cannot excite phonons, and does not contribute to the absorption rate. The second term is the desired one-phonon contribution. The action of the scalar operator on the bound states $|\psi_{nlm}\rangle$ yields
\begin{align}
\langle 0| \frac{\partial \Psi}{\partial x} |\psi_{nlm}\rangle  &= \frac{1}{\sqrt{2 \mu}}\frac{ \partial\psi_{nlm}}{\partial x}
\label{Eqn:FieldAction}
\end{align}
where $|0\rangle$ is the vacuum state and $\psi_{nlm}(x)$ is the spatial wavefunction of $|\psi_{nlm}\rangle$ at $x$.

Using \eqref{Eqn:OperatorExpansion} and \eqref{Eqn:FieldAction} in \eqref{Eqn:MiddleStepThree}, we have
\begin{align}
\langle k' | \Psi \left(x_l\right) |  k, \psi_{nlm}\rangle &\approx \langle k' | \frac{1}{\sqrt{2 \mu}} \left.\frac{ \partial\psi_{nlm}}{\partial x}\right|_{x_l^0} \delta x_l\, |k\rangle
\label{Eqn:TransitionElement}
\end{align}
To evaluate \eqref{Eqn:TransitionElement}, we express $\delta x_l$ in terms of the phonon creation and annihilation operators. Recalling \eqref{Eqn:xjexpansion}, this is
\begin{align}
\delta x_l = \sum_{s = 1}^{N} \frac{y_{ls}}{\sqrt{m_n\omega_s}} (a_s^{\dagger} + a_s)
\end{align}
Inserting this and recalling that $|k'\rangle = |k_1 + 1\rangle \otimes \prod_{s = 2}^{N} |k_s\rangle$, the matrix element evaluates to
\begin{align}
\langle k' | \left( \epsilon\sum_{l = 1}^{N} \Psi \left(x_l\right) \right) |  k, \psi_{nlm}\rangle &\approx \sum_{l = 1}^{N} \frac{\epsilon}{\sqrt{2 \mu}} \left.\frac{ \partial\psi_{nlm}}{\partial x}\right|_{x_l^0} \frac{y_{l1}}{\sqrt{m_n\omega_1}} \sqrt{k_1 + 1}
\label{Eqn:TransitionElementFull}
\end{align}
where $k_1$ is the occupation number of the lowest phonon mode with frequency $\omega_1$ . Suppose the string of neutrons is in equillibrium with a system that has temperature $T$ (in a neutron star, the neutrons are in equllibrium with a gas of electrons in the star, whose temperature ranges between $10^5$ K - $10^9$ K). The occupation number $k_s$ of a mode with frequency $\omega_s$ is
\begin{equation}
k_s  \simeq \frac{T}{\omega_s} \gg 1
\label{Eqn:nsvals}
\end{equation}
For simplicity, let us also convert the sum over the neutron positions in \eqref{Eqn:TransitionElementFull} with an integral over a neutron number density $n(x) \sim N/R$ performed over the stellar medium. With these substitutions and a little
rearranging, \eqref{Eqn:TransitionElementFull} becomes
\begin{align}
\langle k' | \left( \epsilon\sum_{j = 1}^{N} \Psi \left(x_j\right) \right) |  k, \psi_{nlm}\rangle & \approx \frac{\epsilon}{\sqrt{2 \mu} } \sqrt{\frac{T}{\omega_1}} \, \frac{1}{ \sqrt{2m_n \omega_1}}  \int_S dx \, n(x)    \frac{\partial \psi_{nlm}}{\partial x} \, y_1(x)
\label{Eqn:InnerProductFinal}
\end{align}
where $y_1(x)$ is the $s = 1$ wavefunction $y_{1 j}$ written as a function of neutron position $x$ instead of neutron index $j$. With \eqref{Eqn:InnerProductFinal}, we have evaluated the inner products in \eqref{Eqn:MiddleStepThree}. Substituting these results into \eqref{Eqn:ImaginaryPart}, we get the absorption rate
\begin{align}
C_{nlm} \sim \frac{\epsilon^2}{2\mu} \left( \frac{T/\omega_1}{2m_n \omega_1} \right)  \left| \int_S dx\, n(x) \frac{\partial \psi_{nlm}}{\partial x} \, y_1(x) \right|^{2} \left( \frac{\Gamma_1}{(\mu - \omega_1)^2}\right)
\label{Eqn:InnerProductSum}
\end{align}
The integral in the above expression is of course performed only inside the star (of size R).

We now generalize the above computation to three dimensions. In three dimensions, the interaction $\epsilon\, \Psi\, \overline{n}\, n$ can excite phonons in all three directions. The small oscillations of the neutrons about their equillibrium positions can still be diagonalized through transformations similar to \eqref{Eqn:xjexpansion}, where $N$ is now the total number of neutrons in the object. The rest of the calculation goes forward as described in the above paragraphs, with the result
\begin{align}
 C_{nlm} &\sim \frac{\epsilon^2}{2\mu} \left( \frac{T/\omega_1}{2m_n \omega_1} \right)  \left| \int_S d^3\textbf{r}\,  n(r) \nabla\psi_{nlm} \cdot \textbf{y}_1(\textbf{r}) \right|^{2} \left( \frac{\Gamma_1}{(\mu - \omega_1)^2}\right) \label{Eqn:ScalarAbsorptionFull}\\
  &\sim \frac{\epsilon^2}{2\mu} \left( \frac{T/\omega_1}{2m_n \omega_1} \right)  \left| \int_0^R r^2 dr\, n(r) \frac{\partial\psi_{nl}}{\partial r} y_1(r) \right|^{2} \left( \frac{\Gamma_1}{(\mu - \omega_1)^2}\right) \label{Eqn:ScalarAbsorption}
\end{align}
 where the integration is performed inside the star, and in the second step we have estimated the factors in the integral (defining $\psi_{nl}(r) = \psi_{nlm}/Y_{lm}$) for calculational simplicity. This assumes that the phonon wavefunction has the same angular structure as the scalar field, e.g., an $l = m = 2$ scalar excites an $l = m = 2$ phonon. Otherwise, the integral in \eqref{Eqn:ScalarAbsorptionFull} vanishes for a spherical star. (We discuss the impact of deviations from spherical symmetry in Section~\ref{Sec:Stability}.)

The integral in \eqref{Eqn:ScalarAbsorptionFull} also vanishes if the scalar force $\nabla \psi_{nlm}$ is constant. In order to excite a phonon mode in the star, the gradient of the scalar field must change over the extent of the star---a constant force only shifts the center of mass of the star. This condition is satisfied even at lowest order for scalars with $l\neq 1$, and so \eqref{Eqn:ScalarAbsorption} is a good approximation. But we must be more careful with the case $l = 1$. In this case $\nabla \psi_{nlm} = \text{constant}$ at lowest order in $r/a_0$, where $a_0 \gg R$ is the Bohr radius. We must therefore turn to the second-order term for the leading contribution to $C_{nlm}$. This is equivalent to making the substitution $\partial \psi_{nl} / \partial r \to (r / a_0)(\psi_{nl} / \partial r)$ in equation~\eqref{Eqn:ScalarAbsorption} when $l = 1$, and leads to an additional $ \sim (R / a_0)^2$ suppression in $C_{nlm}$. As a result, constraints due to superradiance of the $\psi_{211}$ mode will not be stronger than the constraints due to $\psi_{322}$, despite the larger overlap with the star.
\label{sr:GradientDiscussion}

For masses $\mu$ much bigger than the rotation rate $\Omega$ of the star, the superradiant modes require large $l$. In this case, the high power of $l$ suppresses overlap with the star and thus suppresses the superradiance rate. For $\mu$ much smaller than $R^{-1}$, even though the lowest modes are superradiant, the Bohr radius of the orbit $a_0 = \left(G M \mu^2\right)^{-1}$ is big, leading again to a suppression of the rate. Consequently, as anticipiated in section \ref{Sec:NeutronStarSuperradiance}, superradiance is maximally effective when $\mu \sim \Omega \sim R^{-1}$.

\section{Constraints}
\label{Sec:Constraints}
The absorption coefficients computed in section \ref{Sec:Rate}  can be used to predict the spin down rate of neutron stars. The existence of rapidly rotating pulsars such as PSR J1748-2446ad  \cite{Hessels:2006ze} and PSR B1937+21  \cite{Kulkarni}  can be used to place limits on particles whose existence would have prevented these pulsars from spinning at the observed rates. However, before placing bounds on such particles we first investigate the stability of the superradiant mode. Superradiance can be effective only if there is continuous accumulation of particles into the mode leading to exponential amplification of the mode. If the mode is depleted through some other absorptive process, it will no longer undergo exponential amplification and will not efficiently remove angular momentum from the rotating system. These aspects are discussed in section \ref{Sec:Stability}, where we examine the superradiant instability in realistic astrophysical environments. Following this discussion, we place bounds on possible scalar couplings to neutrons in section \ref{Sec:Results}.

\subsection{Mode Stability}
\label{Sec:Stability}
The modes described by equation \eqref{Eqn:BoundEquation} describe an ideal neutron star with a spherically symmetric mass distribution and an absorption coefficient $C$ that is time independent and constant inside the star. In this ideal world, these modes are eigenfunctions of the Hamiltonian and their growth rate is completely governed by \eqref{Eqn:rate}. However, real neutron stars do not satisfy these conditions. Departures from these symmetries leads to mixing between various modes. In particular, there will be mixing between superradiant and absorptive modes, leading to damping of the superradiant growth. If these mixing terms are appreciable, superradiance will not have a significant impact on the rotational angular momentum of the system.

In the section, we will first describe and develop a formalism to estimate mixing. We will then consider the mixing effects from the free precession of the star, the equatorial bulge in the star caused by rapid rotation, stellar quakes, and tidal disruptions of the system due to companion objects around the pulsar.  We estimate the maximum possible mixing that can be produced in realistic astrophysical situations. This is then incorporated into the parameter space of particle physics models probed by superradiance in Section~\ref{Sec:Results}.

\subsubsection{Overview and Formalism}

The superradiant modes have different azimuthal angular momentum than the absorptive modes. They are therefore mixed together by non-axisymmetric perturbations of the star. Scalars couple to the neutron density and are perturbed by the asymmetries in the mass distribution of the star. Gravitational asymmetries can also cause mixing between modes. These can arise either as a result of asymmetries in the mass distribution of the star or from the presence of companions to the pulsar.

How large a mixing $\delta$ can we tolerate between a superradiant mode  (say $\psi_{l+1, ll}$) and an absorptive mode  (say $\psi_{n'l'm'}$)? In the presence of this mixing,  the physical mode around the star is the linear combination $|\psi_{l+1, ll}\rangle + \delta |\psi_{n'l'm'}\rangle$. The occupation number of this mode changes at a rate $\sim \Gamma_{l+1,ll} + \delta^2 \Gamma_{n'l'm'}$. The mode will grow if this rate is positive, requiring
\begin{equation}
\delta^2 \lesssim -\frac{\Gamma_{l+1,ll}}{\Gamma_{n'l'm'}} \sim \frac{C_{l+1,ll}}{C_{n'l'm'}}
\label{Eqn:GrowthCondition}
\end{equation}
where in the last equality we dropped the kinematic factors that relate the absorption/superradiance rate $\Gamma$ to the non-rotating absorption rate $C$, except for the critical difference in sign.

The most stringent demands on these mixing terms are between that of the superradiant mode $\psi_{l+1,ll}$ and the absorptive modes $\psi_{n00}$, when non-axisymmetries are present to mix those modes. This is due to the fact that the absorption rates  $\Gamma_{nlm}$ are strong functions of the overlap of the mode with the star (see equation \eqref{Eqn:ScalarAbsorption}). The modes $\psi_{l+1,ll}$ carry angular momentum and are localized away from the origin leading to a suppressed overlap with the star. On the other hand, the modes $\psi_{n00}$ do not carry angular momentum and have support at the origin leading to an enhanced absorption rate $\Gamma_{n00}$. Consequently, the superradiance growth condition \eqref{Eqn:GrowthCondition} is the hardest to satisfy for these mixings.

For this paper we restrict our interest to the largest superradiant modes $\psi_{211}$ and $\psi_{322}$, so in this section we will only care to calculate effects that might cause a superradiant mode to mix with absorptive modes that have $l\leq2$. Any modes with higher angular momentum will have a suppressed overlap with the star that would cause them to be absorbed slower than $\psi_{211}$ or $\psi_{322}$ would be superradiantly emitted, even with $\OO(1)$ mixing. We will see below that the allowed mixings are determined by the multipoles of the asymmetries in the system and the usual selection rules. 

In addition to the damping mechanisms introduced by the astrophysical environment, it is theoretically possible that once the particle mode grows, the number density in the mode may become significant enough to cause self interactions that may damp the growth of the mode. Instabilities of this kind were considered in \cite{Arvanitaki:2009fg, Arvanitaki:2010sy} and were not found to be a problem for similar superradiant growth around rotating black hole systems. This is not a surprise since the particles of interest have extremely weak self interaction couplings (such as the QCD axion). This then clears the way to placing limits on various particle physics models in Section~\ref{Sec:Results}.

Before we proceed on to specific sources of mixing, let us briefly develop the general formalism that will provide us with the mixing magnitudes $\delta$. Any non-axisymmetries in the neutron density or gravitational fields will appear as perturbations $H' \propto e^{-i\omega't}$ to the scalar Hamiltonian, and their effect on the Schr\"odinger equation~\eqref{Eqn:BoundEquation} can be estimated using time-dependent perturbation theory. The amplitude of the mixing between initial state $\ket{i}$ and final state $\ket{f}$ with energy difference $\Delta \omega$ is then
\begin{align}
\delta^2 \sim \frac{\left| \bra{f}H'\ket{i} \right|^2}{(\omega' - \Delta \omega)^2}
\label{Eqn:MixingDelta}
\end{align}

To account for mixing due to scalars scattering off neutrons in a non-axisymmetric pulsar, we may perform a calculation very similar to the absorption calculation earlier, this time investigating the real part of the second-order term. In this case, however, we are interested in the elastic scattering process where a scalar $\psi$ is absorbed into a phonon mode, and then re-emitted into a different scalar mode $\psi'$. In a spherically symmetric star, a phonon with wavefunction $y \propto Y_{lm}$ only couples to scalars with $\psi \propto Y_{lm}$. In the presence of a density asymmetry $\delta n\, Y_{LM}$, however, that same phonon can also couple to $\psi' \propto Y_{l\pm L,m+M}$. Let us parameterize the density asymmetry by the amplitude $\delta R$ of the perturbation, such that $\delta n \sim (\delta R / R)\,n$, where $n$ is the average neutron density in the star. Then we can approximate the mixing between scalar modes $\psi$ and $\psi'$ due to some appropriate asymmetry by inserting
\begin{align}
\bra{f}H'_\text{scat}\ket{i} &\sim \frac{\epsilon^2}{2\mu} \left( \frac{T/\omega_1}{2m_n \omega_1} \right)  \left(\frac{\delta R}{R} \right) \left( \int_0^R r^2dr\, n(r) \frac{\partial\psi'^*}{\partial r} y_1(r) \right) \left( \int_0^R r^2dr\, n(r) \frac{\partial\psi}{\partial r} y_1(r) \right) \frac{1}{\mu - \omega_1}
\label{Eqn:PerturbingHamiltonianDensity}
\end{align}
as the matrix element in \eqref{Eqn:MixingDelta}. Note this rate is not suppressed by the decay width $\Gamma_1$, because the scatter is elastic and concerns the real part of the matrix element.

In the sections that follow, we will be considering mixing rates for the two fastest known pulsars PSR~J1748-2446ad (716~Hz) and PSR~B1937+21 (642~Hz). We will use the nominal value $R \sim 12~\textrm{km}$ for both, the measured mass $M = 1.96 M_{\odot}$ for PSR~J1748-2446ad, and the nominal mass $M \sim 1.4M_{\odot}$ for PSR~B1937+21  (see Section~\ref{MassDiscussion} for details).

\subsubsection{Equatorial Bulge and Free Precession}
\label{Sec:Bulge}

Superradiance is effective only in a rapidly rotating neutron star. A rapidly rotating neutron star will not remain spherically symmetric owing to centrifugal pressures that will cause the star to develop an equatorial bulge, giving rise to a quadrupole moment for the star \cite{Laarakkers:1997hb}. But, this rotation by itself does not break the axisymmetry around the rotational axis and hence this quadrupole moment breaks spherical symmetry but preserves axisymmetry. Consequently, this effect leads to mixing between the hydrogenic modes of \eqref{Eqn:BoundEquation} that carry different radial ($n$) and total orbital angular momenta  ($l$) while preserving the azimuthal quantum number $m$, \emph{i.e.} it mixes states of the form $\psi_{nlm}$ and $\psi_{n'l'm}$. Since the azimuthal quantum numbers $m$ are unaffected, this mixing does not couple the superradiant modes with absorptive modes.

However, the rotation axis of a real neutron star will undergo free precession. The rotation axis of the star is tilted from the precession axis by a ``wobble angle'' $\theta_w$, about which it precesses with a frequency $\Omega_p$. These effects break the axisymmetry of the system, leading to coupling between the rotational quadrupole deformation  and modes of different azimuthal angular quantum momenta. Let us first estimate the sizes of these asymmetries before computing their effects on the modes. The free precession frequency $\Omega_p$ of the star is given by $\Omega_p  = \frac{\Delta I}{I} \Omega$ where $I$ is the moment of inertia of the star and $\Delta I$ is its non-axisymmetric piece \cite{Jones:2000iw}. We estimate $\Delta I$ to be of order the quadrupole moment $Q$ induced by the rotation of the star. This has been estimated for a variety of equations of state to be $Q = q G^2 M^3$, with $q \sim 1$ for the most rapidly rotating neutron stars \cite{Laarakkers:1997hb}. Using $Q$, the ratio $\frac{\Delta I}{I} \sim q \left(\frac{G M}{R}\right)^2$, giving rise to a precession frequency $\Omega_p \sim q \left(\frac{G M}{R}\right)^2 \Omega$. Similarly, the maximum ``wobble angle'' $\theta_w$ about which the star can precess without breaking apart has been estimated  to be $\sim 10^{-3} \left(\frac{2\pi \cdot 1\text{kHz}}{\Omega}\right)^2$ \cite{Jones:2000iw}.

We now estimate the mixing that is caused by the wobble $\theta_w$ rotating with a frequency $\Omega_p$. There are two sources that can cause this mixing. First, the gravitational perturbations from the wobble can cause mixing. And secondly, the interaction \eqref{Eqn:ScalarLagrangian} can cause the modes to mix via their interaction with the wobbling stellar medium. To calculate the gravitational effects of the wobble, we must first know the mass moments of the tilted star. We estimate the wobble-induced quadrupole moments by treating the star as a uniform density ellipsoid tilted by a small angle. The resulting moments are given by
\begin{align}
Q_{2m} \sim Q \left(\theta_w Y_{2,1}\,e^{-i \Omega_p t} +  \theta_w^2 Y_{2,2}\,e^{-i 2\Omega_p t} + \text{h.c.} \right)
\end{align}
For a rotating pulsar we have $Q = qG^2 M^3$ as discussed above. Because this wobble induces quadrupole perturbations in the system, it is able to effectively mix the $\psi_{322}$ mode with the strongly absorptive scalar states, such as $\psi_{100}$. It could also mix $\psi_{211}$ with $\psi_{21,-1}$ or $\psi_{210}$, but these three modes have the same overlap with the star and thus comparable superradiance/absorption rates, and we will therefore simply require $\delta^2\lesssim1$.

To understand condition~\eqref{Eqn:GrowthCondition}, we now need to calculate the absorption rate $C_{100}$ of the $\psi_{100}$ mode. This mode couples primarily to the lowest $l = 0$ breathing mode of the star, but this has a frequency similar to the $l = 1$ phonon and $l=2$ phonon that the $\psi_{211}$ and $\psi_{322}$ scalars couple to, respectively \cite{Lindblom:1990}. And, because of the star's rotation (see Section~\ref{GravitationalDampingDiscussion}), the $l=0$ and $l=1$ phonon modes also have similar damping rates, roughly $10^{-1}$ suppressed relative to the $l=2$ phonon. Inserting the hydrogenic wavefunction $\psi_{100}$ into equation~\eqref{Eqn:ScalarAbsorption}, we find the ratios
\begin{align}
\frac{\Gamma_{211}}{\Gamma_{100}} \sim  \frac{\Gamma_{322}}{\Gamma_{100}} \sim 10^{-6} \left( \frac{M}{1.4M_{\odot}} \right)^{2} \left( \frac{R}{12~\textrm{km}} \right)^{2} \left( \frac{\mu}{10^{-11}~\textrm{eV}} \right)^4
\end{align}
With these ratios in hand and an estimate of the mixing from gravitational effects using equations~\eqref{Eqn:MixingDelta} and~\eqref{Eqn:PerturbingHamiltonianDensity}, we find that the condition~\eqref{Eqn:GrowthCondition} is easily satisfied in our region of interest. The wobble-induced gravitational perturbations do not damp the superradiant growth of the scalar modes.

Scattering off neutrons, on the other hand, can provide efficient mixing. The mixing fractions to absorptive modes from scalar-neutron scattering are given by equations~\eqref{Eqn:MixingDelta} and~\eqref{Eqn:PerturbingHamiltonianDensity}. We can estimate the wobble-induced density perturbations by
\begin{align}
\frac{\delta\rho}{\rho_0} \sim
\left( \frac{\Delta I}{I} \right)\left(\theta_w Y_{2,1}\,e^{-i \Omega_p t} +  \theta_w^2 Y_{2,2}\,e^{-i 2\Omega_p t} + \text{h.c.} \right)
\end{align}
Mixing between modes with $\Delta m = |m - m'| = (1~\text{or}~2)$ therefore proceeds with a perturbation of amplitude $\delta R / R \sim (\Delta I / I)\theta_w^{\Delta m} \sim 10^{-7}-10^{-4}$.
Considering the same mixing channels, we find that the $\psi_{211}$ superradiance is not affected, but the $\psi_{322}\to \psi_{100}$ mixing can spoil superradiance of the $\psi_{322}$ mode for large values of the Yukawa coupling. This is folded into our constraint plots.

\subsubsection{Equatorial Ellipticity}
The mass distribution in the star will break axisymmetry at some level. The  multipole moments of this anisotropy will mix modes with different azimuthal angular momenta thereby mixing modes with different azimuthal angular momenta. Distortions from axisymmetry are captured by the dimensionless equatorial ellipticity of the star $\epsilon_s = \frac{I_{xx} - I_{yy}}{I_{zz}}$ \cite{Owen:2005fn} where the $I$s are the moments of inertia of the system about the respective axes. The maximum values of $\epsilon_s$ that can be supported by the star have been estimated to be $\sim 10^{-7}$ \cite{Owen:2005fn}. This asymmetry creates a time dependent perturbation of the star that rotates with the frequency $\Omega$ of the star. Following section \ref{Sec:Bulge}, we estimate that the effects of equatorial ellipiticity are much smaller than those of the free precession of the star. This is because the asymmetry size $\delta R / R \sim \epsilon_s$ of the equatorial ellipiticity is no bigger than the wobble-induced asymmetry, and this perturbation varies at a frequency $\Omega$ larger than the precession frequency $\Omega_p$ responsible for the wobble-induced mixing.

It is also possible that the star may occasionally undergo some internal violent process that causes it to release a sudden burst of radiation. These processes are also not efficient in mixing multiple levels. The change to the total mass of the star caused by such an event is irrelevant since such a change is axisymmetric and cannot mix modes of different azimuthal angular quantum numbers. After the explosion, the equatorial ellipticity of the star will be different than the value it started out with and this change in the ellipticity can mix all the modes. But, the new value of the ellipticity cannot be larger than the maximum allowed by the equation of state of the star. Furthermore, the change to the equatorial ellipticity will also be suppressed by the actual mass lost in the process and since this change must be much less than the actual mass of the star (else the star could not have survived the explosion), the effect of such explosions are insignificant. We treat the effects of ``stellar quakes'' on mixing the modes in the next section.

\subsubsection{Mixing via Phonons}
\label{Sec:PhononMixing}


Stellar quakes may  cause anisotropies in the star and thus produce mixing between superradiant and absorpative modes. Recall from \ref{Sec:Bulge} that the maximal dimensionless ellipticity $\epsilon_s$ that can be supported by the star is roughly $\sim 10^{-7}$. Strictly speaking, this is only a bound on quadrupolar deformations of the star, but we will use it as a proxy to estimate the maximal displacement of any multiple deformation. The pulsar may have undergone violent ``stellar quakes'' in its history, but the displacements caused by such quakes must be smaller than the maximum equatorial ellipticity $\epsilon_s$ that can be supported by the star. We will therefore take $\delta R / R \sim 10^{-7}$ to be a conservative upper bound on the quake-produced phonon amplitudes that might cause mixing. This effect isn't stronger than the wobble-induced mixing for $\psi_{322}$ superradiance, for the same reasons that we can ignore the equatorial ellipticity effect, but for sufficiently large values of the coupling it could serve to spoil $\psi_{211}$ superradiance through mixing with $\psi_{100}$. At worst, this might limit our ability to place constraints above $\epsilon \sim 10^{-18}$, which is already stronger than gravity and so not of great interest to us. 

\subsubsection{Disruptive Companions}
\label{Sec:Companion}
Accretion from the companions is often believed to be the mechanism responsible for powering the initial spin up of the neutron star to the millisecond regime \cite{Kulkarni}, and most millisecond pulsars still have small companions $\lesssim M_{\odot}$ \cite{Kulkarni}. A companion object of mass $M_c$ at a distance $r_c$ will cause tidal disruptions of the bound particle state. The tidal disruption provides dipole and quadrupole gravitational perturbations which can cause the $\psi_{211}$ and $\psi_{322}$ states to get absorbed through mixing with $l = m = 0$ states. Typically we would be most concerned with the $\psi_{100}$ state, since it has the largest absorption rate, but in this case the more dangerous mixing channels are $\psi_{200}$ and $\psi_{300}$ because the smaller energy difference between the initial and final scalars leads to a smaller denominator in the mixing~\eqref{Eqn:MixingDelta}. Expanding the gravitational potential due to the companion at the pulsar, we find the non-zero matrix elements for the desired mixing processes are
\begin{align}
\delta^2_{211\to\textrm{abs}} &\sim \frac{\left| \bra{\psi_{200}}  G \, M_c\,  \mu\, \frac{rY_{1,-1}}{r_c^2}  \ket{\psi_{211}}\right|^2}{(\Omega_c - (E_{200} - E_{211}))^2} \nonumber\\
\delta^2_{322\to\textrm{abs}} &\sim \frac{\left| \bra{\psi_{300}}  G \, M_c\,  \mu\, \frac{r^2Y_{2,-2}}{r_c^3}  \ket{\psi_{322}}\right|^2}{(\Omega_c - (E_{300} - E_{322}))^2}
\label{Eqn:companion}
\end{align}
Unlike the previous mixing processes, where the denominator was always dominated by the oscillation frequency of the perturbation, the denominator in \eqref{Eqn:companion} can be dominated by the energy difference between the states. This is because we wish to describe companions that are relatively far from the star---the time variation $\Omega_c$ from these objects may therefore typically be slower than the energy differences between the states. It will turn out that the orbital rate dominates the denominator for PSR J1748-2446ad, whereas the energy splitting dominates the denominator for the nearly isolated pulsar PSR B1937+21.

The most dangerous mixings are between that of the superradiant mode $\psi_{l+1,ll}$ and the absorptive mode $\psi_{l+1,00}$, instead of $\psi_{100}$ as in the other mixing processes.  This is because the angular frequency of the companion is very low and the energy denominator in \eqref{Eqn:companion} is sensitive to the small energy difference of the states. In Newtonian gravity, these levels are exactly degenerate, up to corrections from deviations from spherical symmetry. This exact degeneracy in Newtonian gravity is a feature of the pure $r^{-1}$ nature of the potential. But, General Relativity induces corrections to this law.  For example, the gravitational effects of angular momentum leads to corrections to the $r^{-1}$ potential, giving rise to familiar effects such as the GR corrections to the precession of the perihelion of Mercury. Similarly, since the states  $\psi_{l+1,ll}$ and $\psi_{l+1,00}$  have different total angular momenta, their energies will also be different. We can estimate this splitting to be roughly $\frac{GM \mu}{r_b} v_{b}^{2} \sim \frac{(GM\mu)^4\mu}{ l^4}$, where $r_b \sim l^2 / (GM\mu^2) $ and $v_b \sim GM\mu / l$ are the radius and typical tangential velocity of the particle's orbit in a Bohr model of this gravitational atom.

The fastest known pulsar PSR J1748-2446ad  with a rotation frequency of 716 Hz \cite{Hessels:2006ze} has a companion of mass $0.1 M_\odot$ at an orbital period $\sim 26$ hours.  The second fastest pulsar PSR B1937+21 (with a rotation frequency 642 Hz) is known to be an isolated pulsar, with an upper bound of $\lessapprox 10^{-9} M_\odot$ on any companion for a distance as large as $\sim 3 \times 10^8$ km \cite{Kulkarni}. Since these are the fastest known pulsars, we will use their existence to impose various bounds on particle physics models in section \ref{Sec:Results}. Inserting these values, we find the condition~\eqref{Eqn:GrowthCondition} prevents superradiant growth for low values of $\mu$ around PSR J1748-2446ad due to its companion, but scalars around the isolated PSR B1937+21 are unaffected by tidal mixing.

Finally, we can also estimate the maximum possible effect of accreting gas on the particle modes. The maximum rate of accretion is limited by the Eddington limit, where the radiation pressure on free electrons balances gravity. This rate is $\sim 4 \times 10^{-8} M_{\odot} \text{yr}^{-1}$ \cite{Kulkarni}.  This estimate is almost certainly an overestimate as the accretion rate should fall as we move away from the star. Using this limit, the maximum amount of mass that could be accreting even out to a distance $r_c \sim 10^7$ km is $\sim 10^{-15} M_{\odot}$, too small to provide any problematic mixing.

In addition to mixing with $\psi_{l+1,00}$, we may also worry about mixing with absorptive modes $\psi_{l+1,l,-l}$.  These have overlap with the star similar to the superradiant modes, leading to absorptive rates $\Gamma_{l+1,l,-l} \sim \Gamma_{l+1,ll}$. Hence, as long as the mixing between these modes is less than 1, the superradiant mode will easily grow. The mixing between them is given by an equation analogous to~\eqref{Eqn:companion}. But, we need to estimate the energy difference between these two states. The GR correction identified in the above paragraph gives an identical contribution to the energies of both states since they have the same total angular momentum. But, since we are dealing with a spinning neutron star, there is an additional contribution to the energies of these states from gravitomagnetism. A spinning object  generates gravitomagnetism  which leads to the analogue of the ``spin-orbit'' coupling between the rotating neutron star and the azimuthal quantum number of the state. This gravitomagnetic field $B_g \sim G \frac{M R^2 \Omega}{r_b^3}$ and it couples to the tangential velocity $v_b \sim \sqrt{\frac{G\,M}{r_b}}$ of the mode. In a mode with non-zero azimuthal angular momentum, $\langle v_b \rangle$ is non-zero and hence this gives rise to an energy splitting $\sim G \frac{M \, \mu \, R^2 \, \Omega}{r_b^2} v_b$. Numerically, we find that this splitting is a tenth or less of the GR correction computed in the above paragraph for the states of interest to us in section \ref{Sec:Constraints}. These mixings will be larger by a factor of 100 or more for the isolated PSR B1937+21, for which the mixing is dominated by the energy splitting. However, since both these states have nearly identifical absorption rates the stability condition \eqref{Eqn:GrowthCondition} is still easily satisfied for mixing between these modes.

\subsection{Results}
\label{Sec:Results}
The estimates in section \eqref{Sec:Stability} suggest that the superradiant mode  can grow in  real astrophysical environments.  The existence of long lived, rapidly rotating pulsars constrains particles that can undergo efficient superradiant growth since superradiant growth occurs at the expense of the rotational energy of the star. We will use the pulsars PSR~J1748-2446ad  (716 Hz) \cite{Hessels:2006ze} and  PSR~B1937+21 (642 Hz) \cite{Kulkarni} to constrain particles that couple to the stellar medium. These pulsars are particularly interesting because not only are they the fastest known pulsars, but their astrophysical environment is also devoid of close, massive companions whose presence may disrupt the growth of the superradiant mode (see section \ref{Sec:Companion}).

The existence of these pulsars implies that the rate \eqref{Eqn:ScalarAbsorption} is small enough so that the pulsars would not have significantly slowed down due to superradiant emission during their lifetime $\tau$. The angular momentum of the star is $L_s \sim 10^{176}\hbar$ and the emission of each particle of mass $\mu$ with azimuthal angular momentum $m \sim 1$ costs angular momentum $ \sim \hbar$. The superradiant mode grows as $e^{\Gamma_{l+1, ll} \tau}$ and we require that this exponential term be smaller than $\sim \frac{L_s}{\hbar}$, implying $\Gamma_{l+1,ll} \lessapprox \frac{176}{\tau}$. Bounds can be placed on particles that fail this test. But, in order to do so, we need to know the age $\tau$ of the pulsar in question.

Reliable upper bounds on the age of the pulsar can be placed from measurements of the spin down rates of the star. The spin down rate gives an estimate of the time required for the frequency of the pulsar to change by an order one amount. This time, called the characteristic age of the pulsar, is $\sim 2 \times 10^{8}$ years for PSR B1937+21 \cite{PhinneyKulkarni}.  Reliable observational lower bounds on the pulsar lifetime are obviously harder to establish. Millisecond pulsars are old objects and are not the result of recent stellar activity \cite{Kulkarni}. In some cases, such as PSR J0034-0534, a lower bound on the age of the pulsar can be determined by observations of the temperature of its companion star \cite{PhinneyKulkarni}, which is correlated with its age. These observations suggest that millisecond pulsars are old objects with ages $\sim 10^{8} - 10^{9}$ years. There are also theoretical arguments that suggest this lifetime. The formation of these rapid pulsars are believed to have been the result of accretion from a nearby companion star. The progenitor neutron star needs to accrete mass $\sim 0.1 M_{\odot}$ in order to achieve the rotation rates observed in milli-second pulsars \cite{Kulkarni}. Accretion at the maximum possible Eddington rate of  $\sim 4 \times 10^{-8} M_{\odot} \text{yr}^{-1}$ \cite{Kulkarni} suggests that the lifetime of the star $\tau$ must be at least $\gtrsim 10^{7}$ years. Consequently, if the accretion proceeds slightly more slowly than the maximum possible Eddington rate, the time neccessary to form the source must be $\gtrapprox 10^{8}$ years. It is thus reasonable to take the age of the pulsar to be equal to the pulsar's measured characteristic age $\sim 10^{8} - 10^{9}$ years. Furthermore, we will suppose that the pulsar has been spinning at its current rate for $\mathcal{O}(1)$ of this lifetime.\label{AgeDiscussion}  

While the characteristic age of PSR B1937+21 has been measured, this determination has not yet been made for PSR J1748-2446ad. Current measurements of the spin down rate of PSR J1748-2446ad suggest a lower bound on its characteristic lifetime $\gtrapprox 2.5 \times 10^7$ years \cite{Hessels:2006ze}. This lower bound is too conservative since formation from accretion would take longer. Instead, we  use the following method to estimate the characteristic age of this object. The pulsar's characteristic age  is determined from its measured rotation rate and the magnitude of its dipole magnetic field. Millisecond pulsars typically have surface dipole magnetic fields clustered around $\sim 3 \times 10^4$ T \cite{PhinneyKulkarni}. Taking this to be the surface magnetic field of PSR J1748-2446ad, we estimate its characteristic age $\sim 10^{9}$ years. With all this in consideration, we conservatively take the stellar lifetime to be $\tau = 3\times10^8$ years for each pulsar in setting our constraints.
Recently \cite{Bassa:2017zpe}, an additional millisecond pulsar, PSR J0952-0607, was disovered. This pulsar has a companion which is 0.02 solar masses (1/10 of J1748's companion) with a period of 6.5 hours (4 times the frequency of J1748's).  Since the matrix element for disruption is linear in mass and quadratic in frequency of the companion, the constraints will be weaker than those of J1748 (in addition, the mass is unknown as of yet).  Thus, we do not include this star on the plots. 

We are nearly ready to place bounds on scalars of mass $\mu$ that couple to neutrons through a Yukawa interaction of strength $\epsilon$.
Stellar parameters such as the temperature $T$, mass (in order to obtain the number of neutrons $N$), radius $R$, the frequency $\omega_1$ of the lowest phonon mode and its damping rate $\Gamma_1$ enter into the estimate of the superradiance rate \eqref{Eqn:ScalarAbsorption}.  For old neutron stars, whose ages are much longer than $10^{6}$ years, the temperature $T$ of the star is $\lessapprox 5 \times 10^5$ K \cite{Yakovlev:2004iq}\label{TemperatureDescription}. However, millisecond pulsars tend to be somewhat warmer, $T \sim 10^7-10^8~\text{K}$ \cite{Gusakov2014}. We will take the pulsar temperatures to be at the lower end of this range, $T \sim 10^7~\text{K}$. The mass of the star is directly obtained from observations of these objects \cite{Hessels:2006ze, Kulkarni} where they exist. The mass of PSR J1748-2446ad has been determined to be $1.96 \pm 0.04 \, M_{\odot}$ \cite{Hessels:2006ze}. The mass of PSR B1937+21 has not been measured and we take it to be equal to the nominal neutron star mass $\sim 1.4 M_{\odot}$  \cite{Kulkarni}. Similarly, the stellar radius for both pulsars is taken to be the nominal size of a neutron star $\sim 12$ km \cite{Kulkarni}. \label{MassDiscussion}

The frequencies and damping rates of phonon modes were estimated in \cite{Lindblom:1990}. In placing bounds we will mostly concentrate on excitations of the lowest-frequency $l=1$ and $l=2$ modes by the absorption of a scalar. We are also interested in the $l = 0$ oscillations for the purposes of mixing estimates (see Section~\ref{Sec:Stability}). In particular, we focus on absorption into the lowest-frequency stellar oscillations, which have 0 radial nodes (or 1, in the case of the dipole oscillation). Of course, one could also include absorption into higher-frequency oscillations with the same angular structure but more radial nodes---but these are at progressively higher frequenices $\omega_n$ and the absorption rate is $\propto \omega_n^{-4}$, so summing over them does not notably enhance the absorption rate. The lowest-frequency $l=0$ and $l=2$ phonons are typically at frequencies $\omega_1 \sim 2 \pi(2~\text{kHz})$, and the $l = 1$ mode is typically at $\omega_1 \sim 2 \pi(4~\text{kHz})$, somewhat higher because the lowest-frequency dipole phonon has a node in the star. Of these, the $l=2$ mode undergoes damping through gravitational radiation with a damping rate $\Gamma \sim 10~\text{Hz}$. The $l=0$ and $l=1$ modes do not damp through gravitational radiation in a non-rotating pulsar. But, in a rapidly rotating pulsar, whose rotational frequency is $\sim$ kHz, these modes will also radiate efficiently through gravitational wave emission, both at a rate suppressed roughly by $ \sim 10^{-1}$ compared to the quadrupole phonon.

We have estimated the $l=0,1$ damping rates by noting that the rotating star has equilibrium density $\rho_0 \propto (Y_{00} + c Y_{20})$, where $c \sim 0.2$ corresponds to the quadrupole moment $Q \sim G^2 M^3$ of a rapidly rotating star \cite{Laarakkers:1997hb}. The continuity equation $\delta \rho = -\nabla \cdot(\rho_0 \boldsymbol{\delta}\textbf{r})$ relates the $\delta r \sim Y_{lm}$ displacements to the resulting density perturbations. Taking simple approximate wave functions for the breathing and dipole phonon modes $\delta r \sim Y_{00}, Y_{11}$, and a stellar density profile $\rho_0 \propto (1-(r/R)^2)$, we find that the $l = 0$ phonon mode decays via quadrupole radiation at a rate $\Gamma_{l = 0} \sim 0.1\, \Gamma_{l = 2}$. The $l = 1$ mode decays via octupole radiation, yet because it oscillates at a higher frequency the decay rate is similar. This is admittedly a rough estimate, but sufficient for us for two reasons: first, both the mixing cutoffs and the $\psi_{211}$ superradiance bounds are only mild functions of $\Gamma$, and second, our strongest bounds in any case come from $\psi_{322}$ superradiance, which is unaffected by these estimates except through the (very mild) effects on mixing. \label{GravitationalDampingDiscussion}  

As we go to higher masses, the superradiant modes will have higher angular momentum. The absorption of these modes will then lead to excitation of phonon modes with $l > 2$. The superradiance rate of these high angular momentum modes is suppressed due to two reasons. First, the overlap of the mode with the star is suppressed, as the modes have high angular momentum. Second, the absorption of these modes results in excitation of modes of high angular momentum in the star. These high angular momentum modes are not as highly damped by gravitational wave emission since they correspond to higher multipole excitations of the star. The damping rates of modes with $l \geq 2$ are given by $\Gamma \sim 10^{5-2l}~\text{Hz}$ \cite{Lindblom:1990}. Both these effects suppress the superradiance rate, limiting the ability of this method to probe scalars of mass $\mu \gg \Omega$. For this reason, we will only place bounds on scalar masses superradiant in the $l = 1$ and $l = 2$ modes.
\begin{figure}[h!]
\begin{center}
\includegraphics[width = 4 in]{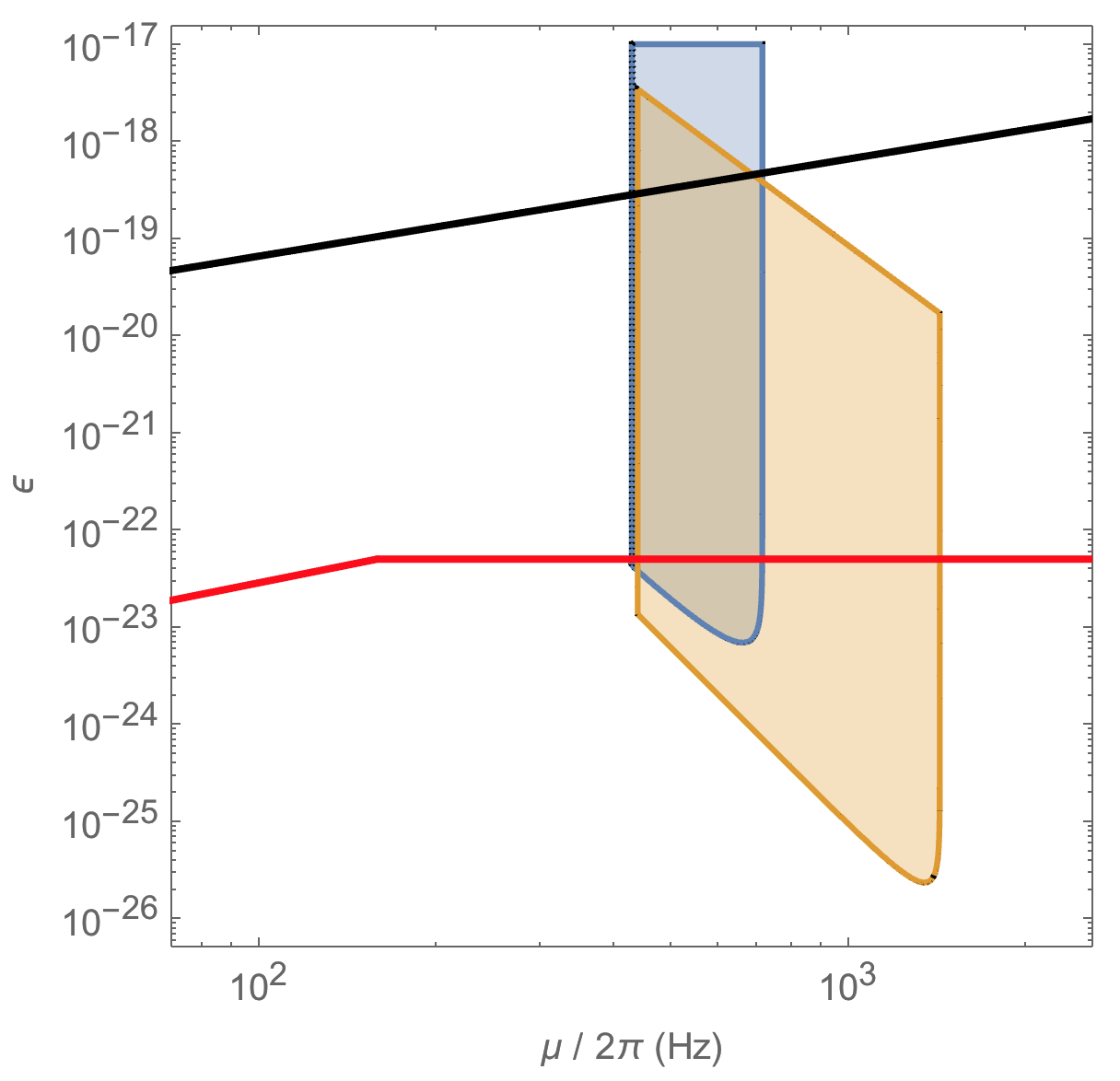}
\caption{ \label{Fig:pulsar716scalar} Constraints imposed by the existence of PSR J1748-2446ad (716~Hz) on scalars of mass $\mu$ with Yukawa coupling $\epsilon$ to neutrons. Shaded regions are excluded due to superradiance into $\psi_{211}$ (blue) and $\psi_{322}$ (orange) scalar modes. The right most boundaries are fixed by the superradiance condition $m\Omega-\mu>0$, and on the left constraints are limited by mixing from companion star tidal forces. The $\psi_{322}$ constraints are limited at large coupling due to mixing through the free precession wobble. The red line shows existing constraints from torsion balance experiments. The black line represents the mass-coupling relation for the QCD axion, assuming $\theta_{eff} \sim 1$ in the star.}
\end{center}
\end{figure}

\begin{figure}
\begin{center}
\includegraphics[width = 4 in]{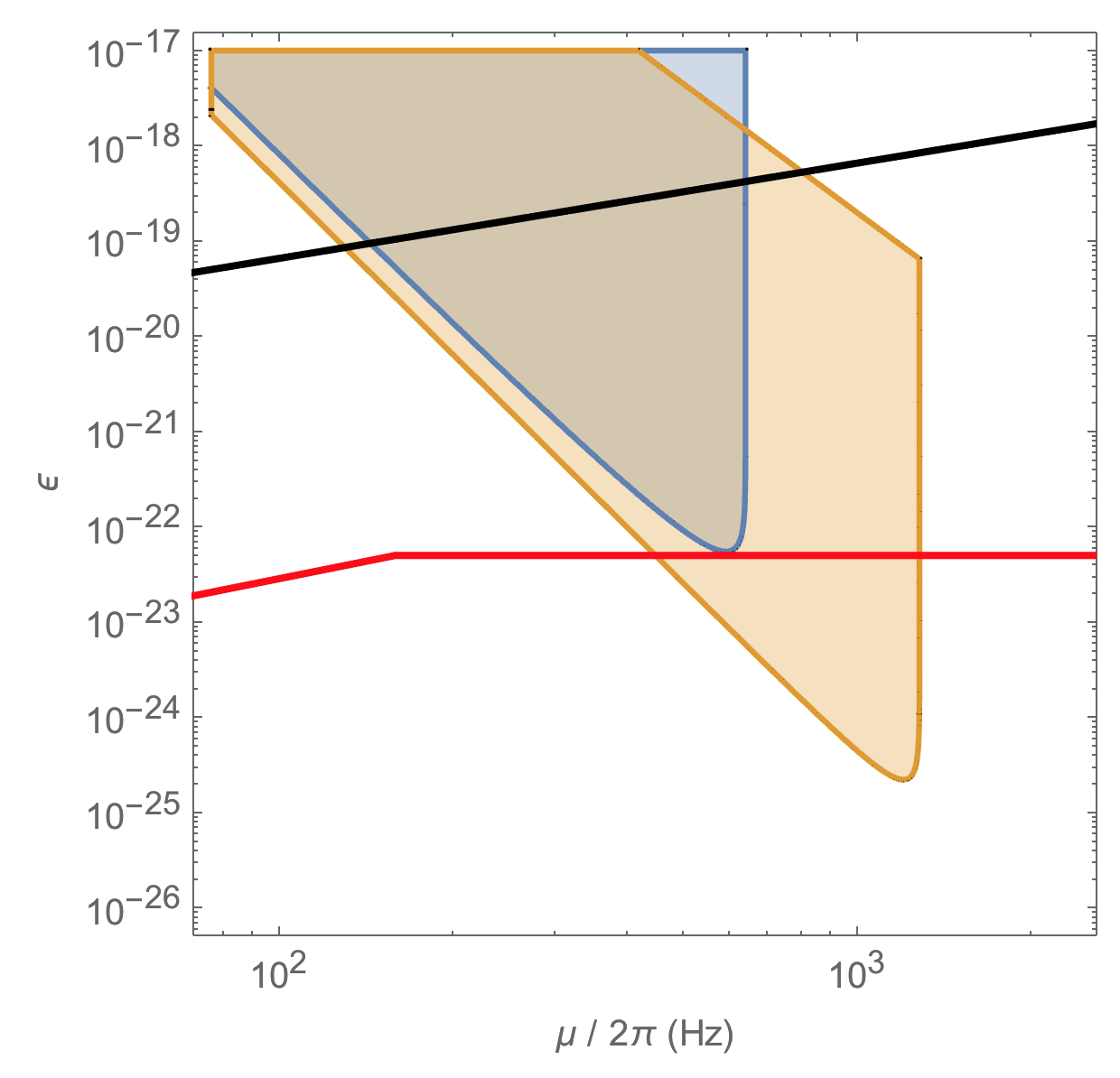}
\caption{ \label{Fig:pulsar642scalar} Constraints imposed by the existence of PSR B1937-21 (642~Hz) on scalars of mass $\mu$ with Yukawa coupling $\epsilon$ to neutrons. Shaded regions are excluded due to superradiance into $\psi_{211}$ (blue) and $\psi_{322}$ (orange) scalar modes. The right most boundaries are fixed by the superradiance condition $m\Omega-\mu>0$. The $\psi_{322}$ constraints are limited at large coupling due to mixing through the free precession wobble. The red line shows existing constraints from torsion balance experiments. The black line represents the mass-coupling relation for the QCD axion, assuming $\theta_{eff} \sim 1$ in the star.}
\end{center}
\end{figure}

\begin{figure}[h!]
\begin{center}
\includegraphics[width = 4 in]{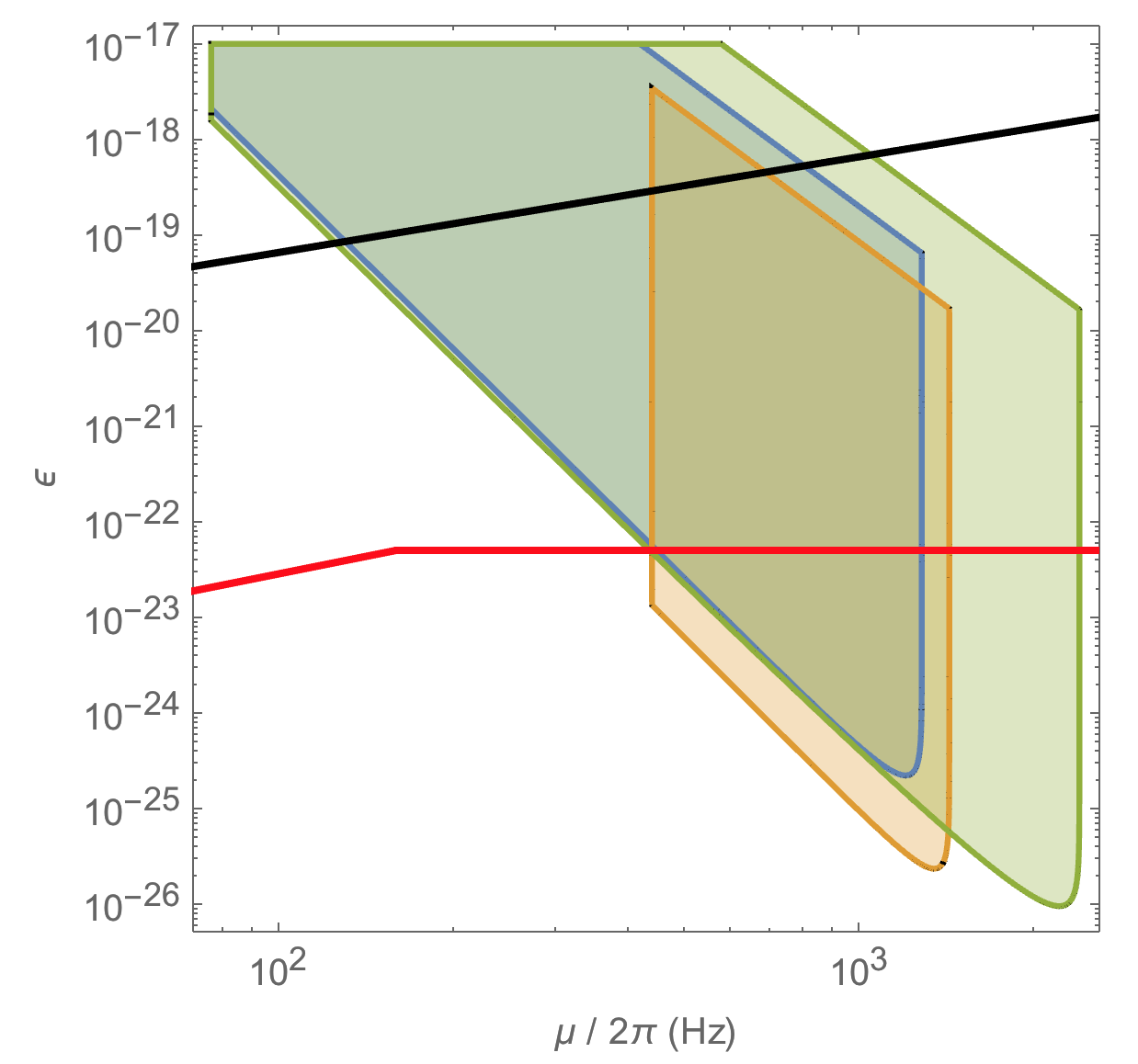}
\caption{ \label{Fig:pulsar1200scalar} Constraints on scalars of mass $\mu$ with Yukawa coupling $\epsilon$ to neutrons. Shaded regions are excluded due to superradiance into $\psi_{322}$ by PSR B1937-21 (642~Hz, blue), PSR J1748-2446ad (716~Hz, orange), and a hypothetical isolated pulsar rotating with a speed of 1200~Hz (green). The PSR J1748-2446ad constraints jut below the others primarily due to the star's larger mass ($1.96~M_{\odot}$ vs $1.4~M_{\odot}$) The red line shows existing constraints from torsion balance experiments. The black line represents the mass-coupling relation for the QCD axion, assuming $\theta_{eff} \sim 1$ in the star.}
\end{center}
\end{figure}

With these parameters, in Figures \ref{Fig:pulsar716scalar} and \ref{Fig:pulsar642scalar},  we place bounds in the $\epsilon - \mu$ plane for scalar interactions with the neutron, using the existence of the pulsars  PSR J1748-2446ad  ($716$ Hz) and  PSR B1937+21 ($642$ Hz) respectively. Figure~\ref{Fig:pulsar1200scalar} represents bounds that could be placed with the discovery of an isolated pulsar rotating at 1200~Hz, and relates them to the other constraints. These bounds consider the superradiant modes $\psi_{211}$ and $\psi_{322}$, coupling respectively to dipolar and quadrupolar oscillations in the star.  We note that existing bounds are $\epsilon \lesssim 5\times10^{-23}$ for most of this parameter space \cite{Adelberger:2003zx}, set by weak equivalence principle tests with torsion balances. We are able to improve on these by up to 3 orders of magnitude.
The bounds are maximally good in the region right near $\mu \sim \Omega$, as expected.

The bounds in Figures \ref{Fig:pulsar716scalar},  \ref{Fig:pulsar642scalar}, and \ref{Fig:pulsar1200scalar} are cut off above and on the left when the superradiant mode is damped by astrophysical anisotropies, primarily the free procession wobble and tidal forces from the companion star (as discussed in section \ref{Sec:Stability}). The upper boundary of the excluded regions are at large couplings when the Yukawa coupling mediates a force comparable to gravity ($\epsilon \sim \sqrt{G} m_n \sim 10^{-19}$).  Here, the free procession wobble causes the $\psi_{322}$ superradiant mode to mix with absorptive modes and wobble-induced scattering to certain absorptive scalar states becomes as efficient as scattering into gravitons (i.e., the superradiant process). 
The constraints due to PSR J1748-2446ad (Figure \ref{Fig:pulsar716scalar}) are additionally limited at low masses $\mu$ due to disruption of the superradiant growth by its companion star, a star of mass $\sim 0.1 M_{\odot}$ at a distance $3.9 \times 10^6$ km away from it. At low masses, the superradiant modes have large Bohr radii with a suppressed overlap with the star, while the damped absorptive modes $\psi_{n00}$ always have support at the origin making their damping rates significantly bigger than the superradiant growth rates. Consequently, the condition \eqref{Eqn:GrowthCondition} becomes increasingly harder to satisfy as tidal forces cause mixing between the superradiant and absorptive modes. PSR~B1937+21 (Figure \ref{Fig:pulsar642scalar}) avoids mixing from a stellar companion because is a nearly isolated pulsar with its closest companion at least $\sim 3 \times 10^{8}$ km away with mass $\lessapprox 10^{-9} M_{\odot}$. We take the hypothetical 1200~Hz pulsar (Figure~\ref{Fig:pulsar1200scalar}) to be similarly isolated. 

Our results take on an additional meaning if indeed $\theta_\textrm{eff} \sim 1$ in a neutron star, as predicted by various neutron star equations of state \cite{Lattimer2001}. In this case, the QCD axion acquires a CP-violating Yukawa coupling to neutrons $\epsilon \sim \theta_\textrm{eff}m_n/f_a \sim m_n/f_a$, and the vertical axis on our plots can be read off as $(f_a / \textrm{GeV})^{-1}$. We are thus able to exclude QCD axions with Planck scale decay constants for specific equations of state of the neutron star.

We have thus constrained any scalars (or pseudoscalars) with a Yukawa coupling \eqref{Eqn:ScalarLagrangian} to neutrons. We improve on the existing torsion balance constraints for scalar masses $2\times 10^{-12}~\text{eV} \lesssim \mu \lesssim 6\times 10^{-12}~\text{eV}$ ($430~\text{Hz} \lesssim \mu/2\pi \lesssim 1420~\text{Hz}$), and (pending the pulsar equation of state) constrain QCD axions with Planck-scale decay constants and masses $5\times 10^{-13}~\text{eV} \lesssim \mu \lesssim 3\times 10^{-12}~\text{eV}$ ($120~\text{Hz} \lesssim \mu/2\pi \lesssim 800~\text{Hz}$).

\section{Conclusions}

The superradiant instabilty of the gravitationally bound states of millisecond pulsars allows their use  as an interesting laboratory to search for light, weakly coupled particles. Measurements from the two fastest known pulsars PSR J1748-2446ad and PSR B1937+21 place bounds on scalars with wavelengths between 100 km  - $10^{4}$ km, improving current bounds by two to four orders of magnitude over this range. Also, if $\theta_\textrm{eff} \sim 1$ in a neutron star as predicted by some equations of state, the QCD axion with a mass in the range $\mu \sim\ $800 Hz - 9000 Hz would be ruled out. The discussions in this paper were restricted to scalars with Yukawa interactions to neutrons. It may be interesting to study a larger class of interacting systems. Using the methods of this paper, it can be readily checked that pseudo-scalar interactions of $\Psi$ with nuclei/electrons cannot be constrained by superradiance using known parameters of milli-second pulsars.
A careful consideration of mixing with absorptive modes in context of the electromagnetic mechanisms of \cite{Cardoso:2017kgn, Day:2019bbh} may allow the results of those papers to be reinterpreted as realistic constraints. Other potentially dissipative mechanisms would also be interesting to investigate, such as an oscillating neutron electric dipole moment induced by an axion-like coupling.

Intriguingly, there appears to be an absence of pulsars with frequencies above $\sim 700$~Hz. This is a puzzling phenomenon since many equations of state of the neutron star can support frequencies well above 1~kHz before break up \cite{Cook:1993qr}. It is unclear if this phenomenon can be explained through standard model processes such as gravitational wave emission, though a variety of astrophysical mechanisms have been proposed \cite{Patruno:2011sj,Bildsten_1998,Cutler:2002nw,Andersson:1998ze,Andersson:1997xt,Lindblom:1998wf}. (Some of these, such as the $r$-mode instability, are superradiance phenomena in their own right.) A particle that is sufficiently strongly coupled to the neutron star medium, with a mass around the breakup frequency, can furnish such a rapid cut off. This explanation could be tested with the discovery of more rapidly rotating pulsars. A pulsar braking mechanism caused by superradiance would lead to the clustering of pulsars at roughly half the mass of the putative particle. A conventional source for damping the stellar rotation such as gravitational wave emission would predict a gentler demise of pulsars on the curve up to rapid rotation. This anomaly may provide an impetus to search for new light particles  that couple to neutrons with mass around $\mu \sim 2\pi \cdot 1500$ Hz in laboratory searches.


\section*{Acknowledgments}

We thank Sergei Dubovsky, Peter W. Graham, Ryan Janish, Lee Lindblom, Riccardo Rattazzi, Oleg Tchernyshyov and Bob Wagoner for useful discussions. DK was supported in part by the NSF under grant PHY-1818899.  SR was supported in part by the NSF under grants PHY-1818899 and PHY-1638509, the Simons Foundation Award 378243 and the Heising-Simons Foundation grant 2015-038.




\end{document}